\documentclass[aps,prl,article,reprint,preprintnumbers,superscriptaddress,noeprint,nofootinbib]{revtex4-1}
\usepackage{amsmath,amssymb}
\usepackage[utf8]{inputenc}
\usepackage{mwe}
\usepackage{xcolor}
\usepackage{comment}
\usepackage[normalem]{ulem}

\usepackage{hyperref}
\hypersetup{colorlinks=true, citecolor=blue, urlcolor=blue, linkcolor=blue}

\newcommand{\ssection}[1]{\emph{#1.---}}

\newcommand{\tint}[0]{\tau^{\text{int}}}
\newcommand{\obs}[0]{\mathcal{O}}

\newcommand{\D}[1]{\mathcal{D}{#1}\;}
\newcommand{\prsample}[0]{U}
\newcommand{\Uone}[0]{\mathrm{U}(1)}

\let\Re\undefined

\DeclareMathOperator{\Re}{Re}

\begin{document}

\title{Equivariant flow-based sampling for lattice gauge theory}
\author{Gurtej~Kanwar}
\affiliation{Center for Theoretical Physics, Massachusetts Institute of Technology, Cambridge, MA 02139, USA}
\author{Michael~S.~Albergo}
\affiliation{Center for Cosmology and Particle Physics,
New York University, New York, NY 10003, USA}
\author{Denis~Boyda}
\affiliation{Center for Theoretical Physics, Massachusetts Institute of Technology, Cambridge, MA 02139, USA}
\author{Kyle~Cranmer}
\affiliation{Center for Cosmology and Particle Physics,
New York University, New York, NY 10003, USA}
\author{Daniel~C.~Hackett}
\affiliation{Center for Theoretical Physics, Massachusetts Institute of Technology, Cambridge, MA 02139, USA}
\author{S\'{e}bastien~Racani\`{e}re}
\affiliation{DeepMind, London, UK}
\author{Danilo~Jimenez~Rezende}
\affiliation{DeepMind, London, UK}
\author{Phiala~E.~Shanahan}
\affiliation{Center for Theoretical Physics, Massachusetts Institute of Technology, Cambridge, MA 02139, USA}

\preprint{MIT-CTP/5181}

\date{\today}
\begin{abstract}
We define a class of machine-learned flow-based sampling algorithms for lattice gauge theories that are gauge-invariant by construction. We demonstrate the application of this framework to U(1) gauge theory in two spacetime dimensions, and find that near critical points in parameter space the approach is orders of magnitude more efficient at sampling topological quantities than more traditional sampling procedures such as Hybrid Monte Carlo and Heat Bath.
\end{abstract}
\maketitle

Many important physical theories are described by Lagrangians that are invariant under local symmetry transformations that form Lie groups; such theories are named {\it gauge theories}.
For example, the Standard Model of particle physics, which is our most accurate description of Nature at the shortest length-scales, is a quantum field theory centered around the action of three gauge groups~\cite{Weinberg:1967,Salam:1968rm,Greenberg:1964,Han:1965}, and several important condensed matter systems can be described by effective gauge theories~\cite{Kogut1979SpinSystems, Seiberg2016GaugeCMTDuality,Sachdev2018EmergentGauge,Guo2020EmergentGauge}. 
In the strong-coupling limit, these theories are non-perturbative, and numerical formulations on discrete spacetime lattices offer the only known way to compute properties of interest from first principles.

Calculations within lattice frameworks typically proceed by estimating expectation values of observables using Markov Chain Monte Carlo (MCMC) to sample from thermodynamic distributions or Euclidean-time path integrals. In both cases, samples $U$ (typically high-dimensional) are drawn from an exponentially weighted distribution $p(U) = e^{-S(U)} / Z$, where the physics is encoded in an energy or action functional $S(U)$, and the normalizing constant $Z$ is unknown. When MCMC sampling from the distribution $p(U)$ is efficient, precise physical predictions can be made from the theory. However, as the model parameters are tuned towards criticality, e.g.~to describe universal properties of condensed matter theories or to access the continuum limit of quantum field theories, critical slowing down (CSD) can cause the computational cost of sampling to diverge~\cite{Wolff:1989wq}.

Specialized approaches have been developed to avoid CSD for specific theories~\cite{Kandel:1988zz,Bonati:2017woi,Wolff:1988uh,Hasenbusch:2017unr,Hasenbusch:2018ztj,Swendsen:1987ce,Prokofev:2001ddj,Kawashima2004,Bietenholz:1995zk,Luscher:2011kk}. For several theories of interest, however, CSD obstructs calculations. This is true in particular for the lattice formulation of quantum chromodynamics (QCD)~\cite{Aoki:2006,Schaefer:2009xx,McGlynn:2014}, which enables calculations of non-perturbative phenomena arising from the Standard Model of particle physics. Recently, there has been progress in the development of \emph{flow-based} generative models which can be trained to directly produce samples from a given probability distribution; early success has been demonstrated in theories of bosonic matter, spin systems, molecular systems, and for Brownian motion~\cite{LiWang2018NNRG,ZhangEWang2018Monge,Albergo2019FlowMCMC,Hartnett:2020,khler2019equivariant,Noeeaaw1147,li2019neural,madhawa2019graphnvp,liu2019graph,shi2020graphaf,both2019temporal}. This progress builds on the great success of flow-based approaches for image, text, and structured object generation~\cite{Dinh:2014,Dinh:2016,Kingma:2018,Ho:2019, tran2019discrete,ziegler2019latent,kumar2019videoflow,gu2019pointflow}, as well as non-flow-based machine learning techniques applied to sampling for physics~\cite{Huang:2017,Wang2017,Wu:2019,zhang2019chem, Carrasquilla_2019}. If flow-based algorithms can be designed and implemented at the scale of state-of-the-art calculations, they would enable efficient sampling in lattice theories that are currently hindered by CSD.

\begin{figure}
\includegraphics{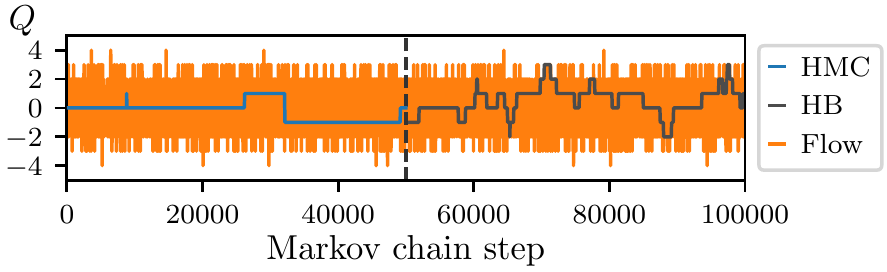}
\caption{Standard approaches (HMC and HB) to MCMC sampling for $\Uone$ gauge theory explore the distribution of topological charge $Q$ very slowly compared with the flow-based approach introduced here. Results are shown for coupling $\beta=7$ on a $16\times 16$ lattice, see Eq.~\eqref{eq:action}. The first (second) half of the Markov chain history is displayed for HMC (HB).}
\label{fig:topo_freezing}
\end{figure}

In this Letter, we develop a provably correct flow-based sampling algorithm designed for lattice gauge theories, including lattice QCD. We demonstrate the application of this approach to $\Uone$ gauge theory in two spacetime dimensions. This theory is solvable, and thus provides a testing ground where the accuracy of numerical methods can be checked. Two standard MCMC approaches, Hybrid Monte Carlo (HMC)~\cite{Duane1987HMC} and Heat Bath (HB)~\cite{Creutz:1979dw,Cabibbo:1982zn,Kennedy:1985nu}, suffer from critical slowing down in this theory; for example, Fig.~\ref{fig:topo_freezing} depicts Markov chain histories for sampling near the continuum limit, in which both methods explore topological sectors very slowly. Using our flow-based algorithm, independent samples of field configurations are produced with appropriate frequency from each topological sector, enabling far more accurate estimation of topological quantities at a given computational cost. Critical to the success of this approach is enforcing exact gauge symmetry in the flow-based distribution: when the symmetry is enforced, we can successfully train flow-based models at a range of parameters approaching criticality, while without it, models of similar scale fail to learn the distributions under the same training procedure.

\ssection{Flow-based sampling}
Flow-based generative models allow sampling from an expressive family of distributions by applying an invertible function $f$ to samples $\prsample$ from a fixed, simple prior distribution defined by a density $r(\prsample)$~\cite{papamakarios2019normalizing}. The resulting samples $U' = f(\prsample)$ are distributed according to a model density $q(U')$. The invertible function is constructed specifically to allow efficient evaluation of the Jacobian factor for any given sample, so that the associated normalized probability density,
\begin{equation} \label{eqn:norm-flow}
  q(U') = q(f(\prsample)) = r(\prsample) \left| \det \frac{\partial f(\prsample)}{\partial \prsample} \right|^{-1},
\end{equation}
is returned with each sample drawn. This feature enables training the flow model, i.e.~optimizing the function $f$, by minimizing the distance between the model probability density $q(U')$ and the desired density $p(U')$ using a chosen metric. Any deviation from the true distribution due to an imperfect model can be corrected by a number of techniques; in this work, we apply Metropolis independence sampling~\cite{Albergo2019FlowMCMC}.\footnote{Reweighted observables can also be used~\cite{Ferrenberg:1988yz,mller2018neural}. This is efficient when measurements of the action are more costly than measurements of observables.}

A powerful approach to defining a flexible invertible function $f$ is through composition of several \emph{coupling layers}, $f := g_m \circ \dots \circ g_1$. Coupling layers act on samples $U$ by applying an analytically invertible transformation (such as a scaling) to a subset of the components $U^{A} := \left\{ U^{i} : i \in A \right\}$, where the superscript $i$ indexes components of the multi-dimensional sample $U$ and the set $A$ indicates the components that are transformed. The remaining (unmodified) components $U^B$, defined by $U^B = U \setminus U^A$, are given as input to a feed-forward neural network that parameterizes the transformation. This variable splitting guarantees invertibility despite the use of non-invertible feed-forward networks.

\ssection{Gauge-invariant flows}
Lattice gauge theories can be defined in terms of one gauge variable $U_\mu(x)$ per nearest-neighbor link $(x,x+\hat{\mu})$ of the lattice. Samples thus live in the compact manifold $G^{N_d V}$, where $G$ is the manifold of the gauge group, $N_d$ is the spacetime dimension, and $V$ is the lattice volume. The physical distribution $p(U)$ is exactly invariant under a discrete translational symmetry group with $V$ elements and a continuous $V$-dimensional gauge symmetry group, meaning that the density associated with any transformed field configuration $\widetilde{U}$ is identical to that of the untransformed configuration, $p(\widetilde{U}) = p(U)$. Under a \emph{gauge transformation}, links $U_\mu(x)$ are transformed by a group-valued field $\Omega(x)$ as
\begin{equation} \label{eqn:gauge-transform}
  U_{\mu}(x) \; \rightarrow \; \widetilde{U}_\mu(x) = \Omega(x) U_{\mu}(x) \Omega^\dagger (x+\hat{\mu}).
\end{equation}

In the flow-based approach, symmetries correspond to flat directions of the probability density that must be reproduced by the model. Exactly encoding symmetries in machine learning models can improve training and model quality compared with learning the symmetries over the course of training~\citep{CohenWelling2016,Cohen:2019,khler2019equivariant,rezende2019equivariant,wirnsberger2020targeted, ZhangEWang2018Monge,Finzi:2020}. The incorporation of translational symmetries into models is possible using convolutional architectures, as studied for example in Ref.~\cite{CohenWelling2016}. To address gauge symmetry, one could attempt to use a gauge-fixing procedure to select a single configuration from each gauge-equivalent class, leaving only physical degrees of freedom; however, the only known gauge fixing procedures that preserve translational invariance are based on implicit differential equation constraints~\cite{DeGrand:2006zz}, which do not have a straightforward implementation in flows. Here, we instead introduce a method to preserve exact gauge invariance in flow-based models.

When a flow-based model is defined in terms of coupling layers, its output distribution will be invariant under a symmetry group if two conditions are met:
\begin{enumerate}
    \item The prior distribution is symmetric.
    \item Each coupling layer is \emph{equivariant} under the symmetry, i.e.~all transformations commute through application of the coupling layer~\cite{Hinton:2011,kivinen2011transformation,schmidt2012learning,CohenWelling2016,rezende2019equivariant}.
\end{enumerate}
Choosing a prior distribution that is symmetric is typically straightforward, for example a uniform distribution with respect to the Haar measure over gauge links is both translationally and gauge invariant. Using gauge-equivariant coupling layers with such a prior distribution then defines a gauge-invariant flow-based model.

\ssection{Gauge-equivariant coupling layers}
We construct an explicitly gauge-equivariant and invertible coupling layer $g: G^{N_d V} \rightarrow G^{N_d V}$ by splitting the input variables into subsets $U^A$ and $U^B$. In terms of these subsets, we define the action of the coupling layer to be ${g(U^A, U^B) = ({U'}^A, U^B)}$, where link $U^i \in U^A$ is mapped to
\begin{equation} \label{eqn:gauge-equiv-coupling}
{U'}^i = h(U^i S^i | I^i) {S^i}^{\dagger},
\end{equation}
in which $h: G \rightarrow G$ is an invertible \emph{kernel} which is explicitly parameterized by a set of gauge-invariant quantities $I^i$ constructed from the elements of $U^B$. Here, $S^i$ is a product of links such that $U^i S^i$ forms a loop that starts and ends at a common point $x$, and therefore transforms under the gauge symmetry to $\Omega(x) \, U^i \Omega^\dagger(x+\hat{\mu}) \, \Omega(x+\hat{\mu}) S^i \Omega^\dagger(x) = \Omega(x) \, U^i S^i \, \Omega^\dagger(x)$. With this definition, the coupling layer is gauge equivariant if the kernel satisfies
\begin{equation} \label{eqn:kernel-transform}
    h(X W X^\dagger) = X \, h(W) \, X^\dagger, \quad \forall X, W \in G,
\end{equation}
which implies that ${U'}^i \rightarrow {\widetilde{U}}^{\prime i}$ transforms according to Eq.~\eqref{eqn:gauge-transform},
\begin{equation}
\begin{aligned}
    {\widetilde{U}}^{\prime i} &= h\left( \Omega(x) \, U^i S^i \, \Omega^\dagger(x) | I^i \right) \; \Omega(x) {S^i}^\dagger \Omega^\dagger(x+\hat{\mu}) \\
    &= \Omega(x) {U'}^i \Omega^\dagger(x+\hat{\mu}).
\end{aligned}
\end{equation}
To ensure invertibility, the product of links $S^i$ must not contain any links in $U^A$.

For an Abelian group, the transformation property in Eq.~\eqref{eqn:kernel-transform} is trivially satisfied by any kernel. In the $\Uone$ gauge theory considered below, we therefore define the kernel using invertible flows parameterized by neural networks. For non-Abelian theories, it has been shown that it is possible to construct invertible functions on spheres~\cite{rezende2020normalizing} and surjective functions on general Lie groups~\cite{falorsi19lie}. If these approaches can be generalized to produce invertible functions with convergent power expansions, they will satisfy the necessary kernel transformation property, since ${h(X W X^\dagger) = \sum_n \alpha_n (X W X^\dagger)^n = X \, h(W) \, X^\dagger}$.

\begin{figure}
    \centering
    \includegraphics[width=\columnwidth,trim=0 0 0 25, clip]{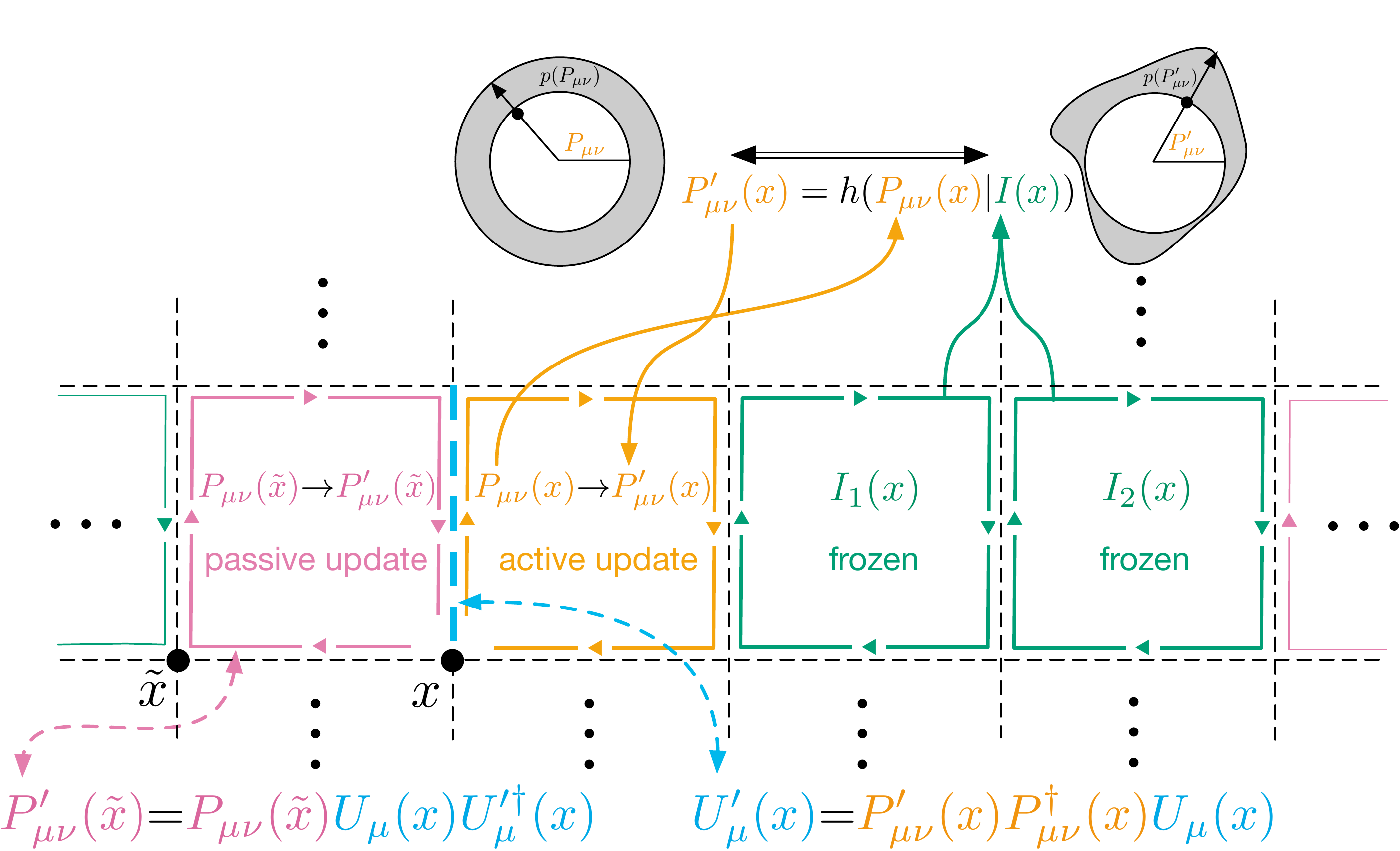}

    \caption{An example of a variable splitting based on a tiled $4\times 1$ pattern with an actively updated link $U^i \equiv U_\mu(x)$ and $1\times1$ loop $U^i S^i \equiv P_{\mu\nu}(x)$ located at $x$, a passively updated $1\times1$ loop at $\widetilde{x}=x-\hat{\nu}$, and two frozen traced $1\times1$ loops at $x+\hat\nu$ and $x+2\hat\nu$ included in the set $I^i$.
    }
    \label{fig:equiv_masking}
\end{figure}

An example of a variable splitting suitable for both Abelian and non-Abelian gauge theories is given by the pattern depicted in Fig.~\ref{fig:equiv_masking}. In this example, the set of updated links $U^A$ consists of vertical links spaced by 4 sites, and the products $U^i S^i$ are $1\times1$ loops adjacent to each $U^i$. This is sufficiently sparse such that every $S^i$ is independent of all updated links in $U^A$, and a non-trivial set of invariants $I^i$ (e.g.~all traced $1\times1$ loops that are not adjacent to updated links) can be constructed to parametrize the transformation. Composing coupling layers using rotations and offsets of the pattern allows all links to be updated.\footnote{For example, composing 8 such layers is sufficient to update all links in $2D$.}

Using gauge-equivariant coupling layers constructed in terms of kernels generalizes the ``trivializing map'' proposed in Ref.~\cite{Luscher:2009eq}. There, repeatedly applying a specific kernel based on gradients of the action theoretically trivializes a gauge theory, i.e.~maps the Euclidean time distribution to a uniform one. The family of gauge equivariant flows defined here includes the trivializing map in the limit of a large number of coupling layers and arbitrarily expressive kernel, indicating that in this limiting case exact sampling as described in Ref.~\cite{Luscher:2009eq} is possible. However, the approach presented here allows for more general and inexpensive parametrizations of $h$. These can be optimized to produce flows that similarly trivialize the theory, and which may have a lower cost of evaluation than implementations of the analytical trivializing map~\cite{Engel2011TestTrivMap}.

\ssection{Application to U(1) gauge theory}
Gauge theory with a $\Uone$ gauge group defined in two spacetime dimensions is the quenched limit of $1+1D$ electrodynamics, i.e.~the Schwinger model~\cite{Schwinger:1962tp}. The full Schwinger model reproduces many features of quantum chromodynamics (confinement, an axial anomaly, topology, and chiral symmetry breaking) while being analytically tractable. Even in the quenched limit, the well-defined gauge field topology results in severe critical slowing down of MCMC methods for sampling lattice discretizations of the model as the coupling is taken to criticality. We consider the lattice discretization given by the Wilson gauge action~\cite{Wilson:1974sk},
\begin{equation}\label{eq:action}
    S(U) := -\beta \sum_x \Re P(x),
\end{equation}
where $P(x)$ is the plaquette at $x$ defined in terms of link variables $U_\mu(x) \in \Uone$,
\begin{equation} \label{eq:plaquette}
\quad P(x) := U_{0}(x) U_1(x+\hat{0}) U^\dagger_{0}(x+\hat{1}) U^\dagger_{1}(x),
\end{equation}
and $x = (x_0, x_1)$ runs over coordinates in an $L \times L$ square lattice with periodic boundary conditions. Physical information may be extracted from the model by considering expectation values of observables $\obs$ under the Euclidean time path integral,
\begin{equation}
  \left< \obs \right> := \frac{1}{Z} \int \D{U} \obs(U) e^{-S(U)},
\end{equation}
where $\int \D{U}$ denotes integration over the product of Haar measures for each link, and $Z=\int\D{U} e^{-S(U)}$.
In this study, three key observables were considered:
\begin{enumerate}
    
    \item Expectation values of powers of plaquettes.
    
    \item Expectation values of $\ell \times \ell$ Wilson loops $W_{\ell \times \ell} = \prod_{x \in \ell \times \ell} P(x)$.
    
    \item Topological susceptibility $\chi_Q = \langle Q^2 / V \rangle$, where topological charge $Q := \frac{1}{2\pi} \sum_x \arg\left(P(x)\right)$ is defined in terms of plaquette phase in the principal interval, $\arg\left(P(x)\right) \in [-\pi, \pi]$.
    
\end{enumerate}

To investigate critical slowing down, we studied the theory at a fixed lattice size, $L = 16$, using seven choices of the parameter $\beta = \{ 1,2,3,4,5,6,7 \}$; the theory approaches the continuum limit as $\beta \rightarrow \infty$. For each parameter choice, we trained gauge invariant flow-based models using a uniform prior distribution and a composition of 24 gauge-equivariant coupling layers. The kernels $h$ were chosen to be mixtures of Non-Compact Projections~\cite{rezende2020normalizing}, which are suitable for $\Uone$ group elements; in particular, we used 6 components for each mixture and parameterized each transformation with a convolutional neural network. The model architecture was held fixed across all choices of $\beta$, ensuring identical cost to draw samples for each parameter choice. To train the models, we minimized the Kullback-Leibler divergence between the model density $q(U)$ and the target density $e^{-S(U)} / Z$. Training was halted when the loss function reached a plateau. For this proof-of-principle study, we did not perform extensive optimization over the variable splitting pattern, neural network architecture, or training hyperparameters, and it is likely that better models can be trained.

After training, the flow-based models were used to generate proposals for a Metropolis independence Markov chain~\cite{Albergo2019FlowMCMC}, producing ensembles of $100,000$ samples each.
For comparison, ensembles of identical size were produced using the HMC and Heat Bath algorithms. For all choices of $\beta$, we fixed the HMC trajectory length to achieve $> 80\%$ acceptance rate when using a leapfrog integrator with $5$ steps. Each HB step was defined as one sweep, i.e.~a single update of every link.
To within $10\%$, the computational cost per HMC trajectory was equal to the cost per proposal from the flow-based model in a single-threaded CPU environment, while the cost per Heat Bath step was half that of HMC or flow.

\begin{figure}
    \centering
    \includegraphics{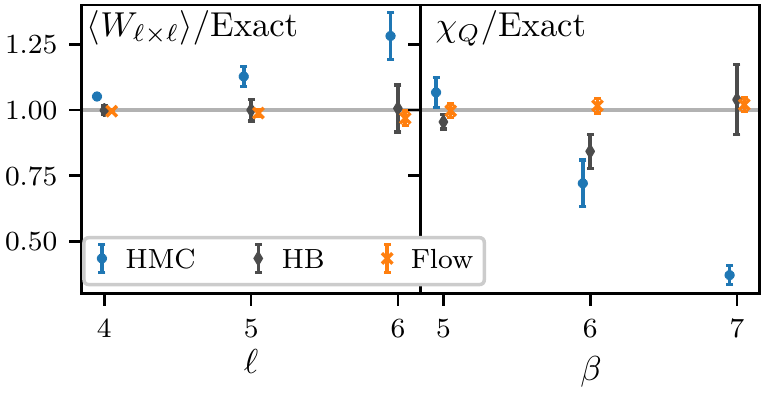}
    \caption{Left: estimates of average Wilson loops $\langle W_{\ell \times \ell} \rangle$ measured on the finest ensemble studied here ($\beta = 7$). Right: estimates of topological susceptibility measured on the three finest ensembles studied here ($\beta = 5,6,7$).  All values are plotted as ratios to the exact results. The flow-based estimates are consistent with the exact values, while the HMC and Heat Bath estimates have larger uncertainties and also significantly deviate from the exact values in some cases.
    }
    \label{fig:compare_obs}
\end{figure}

Using samples from a flow-based model as proposals within a Markov chain ensures unbiased estimates after thermalization; at the finite ensemble size used here, all observables were found to agree with analytical results within statistical uncertainties. Of the observables we studied, local quantities like powers of plaquettes and expectation values of small Wilson loops were estimated more precisely by HMC and HB than with the flow-based algorithm. However, Fig.~\ref{fig:compare_obs} shows that for observables with larger extent such as $W_{\ell \times \ell}$ with $\ell \geq 4$, and particularly for $\chi_Q$, large autocorrelations in the HMC and HB samples result in estimates that deviate from the exact values and have lower precision than the flow-based estimates.

\begin{figure}
    \centering
    \includegraphics{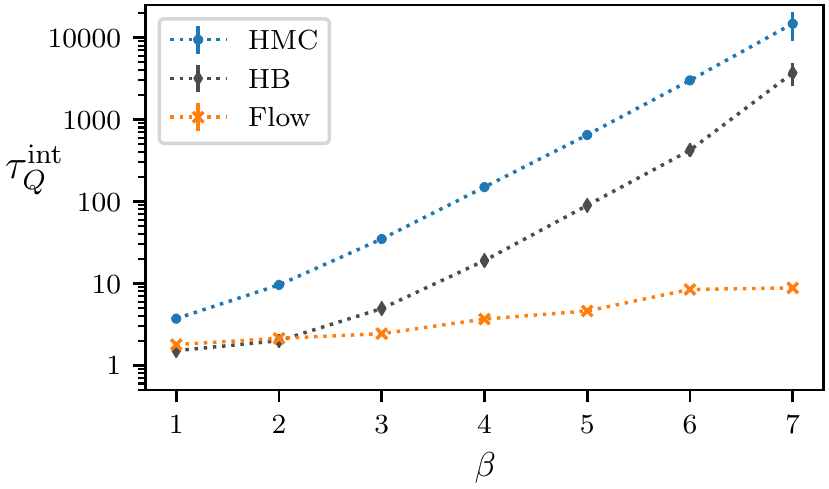}
    \caption{Integrated autocorrelation time for the topological charge, $\tint_Q$, measured on ensembles of $16 \times 16$ lattices generated using HMC, Heat Bath, and the flow-based algorithm. Ten replicas of each ensemble were used to estimate errors, which are smaller than the plot markers for most points.}
    \label{fig:tint_hmc_vs_flow}
\end{figure}

For Markov chain methods, the characteristic length of autocorrelations for an observable $\obs$ can be defined by the integrated autocorrelation time $\tint_\obs$~\cite{Madras:1988ei}. Fig.~\ref{fig:tint_hmc_vs_flow} compares $\tint_Q$ for HMC and HB to that in the flow-based algorithm as an indicator of how well the three methods explore the distribution of topological charge. For all three methods, $\tint_Q$ grows as $\beta$ is increased towards the continuum limit. However, this problem is far less severe for the flow-based algorithm than for HMC or HB. For example, the autocorrelation time in the flow-based algorithm is approximately $10$ at the largest value of $\beta$, whereas $\tint_Q \approx 4000$ for HB and $\tint_Q \approx 15000$ for HMC. Accounting for the relative cost per step of each Markov chain, the flow-based Metropolis sampler is therefore roughly $1500$ times more efficient than HMC and $200$ times more efficient than Heat Bath in determining topological quantities. A promising possibility for further development is mixing flow-based Markov chain steps with HMC trajectories or Heat Bath sweeps to gain the benefits of standard Markov chain steps for local observables and of the flow-based algorithm for extended and topological observables.

\ssection{Summary}
Critical slowing down of sampling in lattice gauge theories is an obstacle to precisely estimating quantities of physical interest as critical limits of the theories are approached. Flow-based models enable direct sampling from an approximation to the distribution of interest, from which estimates of physical observables can be derived that are exact in the infinite-statistics limit. Here we introduce flow-based models constructed to satisfy exact gauge invariance, and find that applying this approach to a two-dimensional Abelian gauge theory enables more efficient estimation of topological quantities than existing algorithms such as HMC and Heat Bath.

The approach presented here is generally applicable to gauge theories defined by Lie groups, including non-Abelian theories such as QCD. To extend this method to such theories, expressive invertible functions must be defined as the kernels of gauge equivariant coupling layers. There are several possible avenues forward; for example, Ref.~\cite{rezende2020normalizing} defines flows on spherical manifolds and Ref.~\cite{falorsi19lie} defines surjective (though not bijective) maps on Lie groups, both using neural network parameterizations. Future work will explore constructing kernels based on generalizations of these methods, and thus producing gauge invariant flows for non-Abelian theories like QCD.

\begin{acknowledgments}
\ssection{Acknowledgments}
The authors thank Jiunn-Wei Chen, Will Detmold, and Andrew Pochinsky for helpful discussions.
GK, DB, DCH, and PES are supported in part by the U.S.~Department of Energy, Office of Science, Office of Nuclear Physics, under grant Contract Number DE-SC0011090. PES is additionally supported by the National Science Foundation under CAREER Award 1841699.
KC is supported by the National Science Foundation under the awards ACI- 1450310, OAC-1836650, and OAC-1841471 and by the Moore-Sloan data science environment at NYU. MSA is supported by the Carl Feinberg Fellowship in Theoretical Physics. This work is associated with an ALCF Aurora Early Science Program project, and used resources of the Argonne Leadership Computing Facility, which is a DOE Office of Science User Facility supported under Contract DE-AC02-06CH11357.
\end{acknowledgments}

\bibliographystyle{apsrev4-1}
\bibliography{main}

\end{document}